\documentclass[12pt, onecolumn, draftcls]{IEEEtran}

\newcommand{\qed}{\mbox{}\hspace*{\fill}\nolinebreak\mbox{$\rule{0.55em}{0.55em}$}} 

\newcommand{\paren}[1]{\left(#1\right)}
\newcommand{\sqparen}[1]{\left[#1\right]}
\newcommand{\brparen}[1]{\left\{#1\right\}}
\newcommand{\field}[1]{\ensuremath{\mathbb{#1}}}
\newcommand{\R}{\ensuremath{\field{R}}} 
\newcommand{\Rp}{\ensuremath{\R_+}} 
\newcommand{\1}{\ensuremath{\mathbf{1}}} 
\newcommand{\I}[1]{\ensuremath{\mathsf{1}_{\left\{#1\right\}}}} 
\newcommand{\EW}{\ensuremath{\mathsf{E}}} 
\newcommand{\ES}[1]{\ensuremath{\mathsf{E}\left[#1 \right]}} 
\newcommand{\e}[1]{\ensuremath{{\rm e}^{#1}}} 

\newcommand{\SINR}{\ensuremath{{\rm SINR}}}
\newcommand{\SNR}{\ensuremath{{\rm SNR}}}

\newcommand{\BO}[1]{\ensuremath{O\paren{#1}}}

\newcommand{\pc}{\ensuremath{\vec{P}}}

\newcommand{\insrate}[2]{\ensuremath{R_{#1}\paren{#2}}}
\renewcommand{\vec}[1]{\ensuremath{\mathbf{#1}}}
\newcommand{\imax}[2]{\ensuremath{#1}_{\ensuremath{{\paren{#2}}}}}
\newcommand{\major}{\ensuremath{\succeq_{\rm M}}}
\newcommand{\dom}{\ensuremath{\mathcal{D}}}
\newcommand{\diag}[1]{\ensuremath{{\rm \bf diag}\paren{#1}}}

\newtheorem{theorem}{Theorem}
\newtheorem{lemma}{Lemma}

\newtheorem{example}{Example}
\usepackage{amsmath, latexsym, amsfonts, amssymb}
\usepackage[mathscr]{eucal}
\usepackage{graphicx}
\usepackage{array, tabularx, colortbl, color, threeparttable}
\usepackage[table]{xcolor}
\usepackage{psfrag}
\usepackage{multirow}
\usepackage{booktabs}
\usepackage{algorithm}
\usepackage{algorithmic}

\usepackage{authblk}

\begin{document}
\title{Optimality of binary power-control in a single cell via majorization \thanks{This research was supported in part by the Australian Research Council, under Grant DP-11-0102729, and NUS startup grant 263-000-572-133. The authors can be contacted at email addresses: hazeri@unimelb.edu.au and elehsv@nus.edu.sg, respectively.}}


\author[1]{Hazer Inaltekin}
\author[2]{Stephen V. Hanly}
\affil[1]{{\small Department of Electrical and Electronic Engineering,
University of Melbourne, Australia.}}
\affil[2]{{\small Department of Electrical and Computer Engineering,
National University of Singapore, Singapore.}}

\date{}
\maketitle
\thispagestyle{empty}

\begin{abstract}
\textbf{This paper
considers the optimum single cell power-control maximizing the
aggregate (uplink) communication rate of the cell when there are
peak power constraints at mobile users, and a low-complexity data
decoder (without successive decoding) at the base station. It is
shown, via the theory of majorization,  that the optimum power
allocation is {\em binary}, which means links are either ``on'' or
``off''.  By exploiting further structure of the optimum binary
power allocation, a simple polynomial-time algorithm for finding the
optimum transmission power allocation is proposed, together with a
reduced complexity near-optimal heuristic algorithm. Sufficient
conditions under which channel-state aware
time-division-multiple-access (TDMA) maximizes the aggregate
communication rate are established.  Finally, a numerical study is performed to compare and contrast the
performance achieved by the optimum binary power-control policy with
other sub-optimum policies and the throughput capacity achievable
via successive decoding.  It is observed that two
dominant modes of communication arise, wideband or TDMA, and that
successive decoding achieves better sum-rates only under near
perfect interference cancellation efficiency.}
\end{abstract}
\begin{keywords}
Communication networks, power-control, sum-rate capacity, TDMA, successive decoding
\end{keywords}
\section{Introduction}

\PARstart{N}{ext} generation 4G wireless communication systems are
required to support all-IP services including high data rate
multimedia traffic \cite{WSJ10}, with bit rate targets as high as
$1$ Gbit/s for low mobility, and $100$ Mbit/s for high mobility
\cite{ITU08}. Transmission at such high rates is certainly
achievable today on point-to-point links, using the great advances
made in wireless communications over the past couple of decades. But
in wireless networks, including 4G systems, interference between
links remains as a fundamental bottleneck that needs to be overcome
\cite{Tse09}. Part of the challenge arises from the broadcast nature
of the shared wireless medium: transmission power has to be
allocated to each link, but this allocation has knock-on effects on
other links in the network. Much progress has been made on this
problem when target rates are specified for each user and the
objective is to minimize total transmit power in the network
\cite{HT99}. However, solving for optimum power allocations that
maximize the total Shannon-theoretic sum-rate in the presence of
interfering links seems to be much harder: It is generally a {\em
non-linear}, {\em non-convex} constrained optimization problem
\cite{Zhang08}. This motivates a search for structure leading to
simplifications in the power allocation problem for sum-rate
maximization.

In this paper, we focus on the optimum allocation of transmission
powers to mobile terminals in order to maximize the total
communication sum-rate when a low-complexity single-user decoder
(without successive decoding) is used at the base station. This is
the conventional single cell matched filter detection based uplink model: All mobiles are in the same
cell and must all be decoded at the same base station. Even though
this optimization problem is non-convex, we solve it by identifying
an underlying {\em Schur-convex} structure in the objective sum-rate
function.  We show that the optimum power allocation is {\em
binary}, {\em i.e.,} a user either transmits with full power or does not
transmit at all. By utilizing the binary structure of the sum-rate
maximizing optimum power allocation, we observe two dominant modes
of communication: either the best user transmits with full power,
which can be considered a channel quality based
time-division-multiple-access (TDMA) mode, or all users transmit
with full power, which can be considered a wideband (WB) mode.  This
result has implications for implementing joint power-control and
scheduling, and helps to theoretically justify existing engineering
approaches, such as code-division-multiple-access (CDMA), and scheduling based on channel quality.

We also compare sum-rates achieved by the optimum power-control
policy with throughput capacity limits that can be achieved by
successive decoding.  Our results indicate that gains over the
simple optimum binary power-control due to advanced interference
cancellation techniques can be harvested only if the cancellation
efficiency is near-perfect.

\section{Related work}

In this paper, we are motivated by recent work on interference networks that shows that binary power-control is often close to
optimal when interference is treated as Gaussian noise, links have maximum (peak) power constraints, and the objective is to maximize the sum-rate, even if it is not necessarily optimal in general \cite{Gesbert08}.  ``Binary'' here just means that a link is either ``on'' or ``off'', either at zero power, or maximum power, without taking any value in the continuum of possible values between $0$ and the peak power level.

In addition to \cite{Gesbert08}, some other works such as  \cite{JG03}, \cite{OZW03} and \cite{Oh06} also motivate us to investigate the optimality of binary power-control.  Both \cite{JG03} and \cite{OZW03} consider jointly optimal allocation of rates and transmission powers in CDMA networks under alternative objectives such as maximization of the sum of signal-to-interference-plus-noise-ratios ($\SINR$) \cite{JG03} and the packet success probability \cite{OZW03}.  Both approaches convert the problem into a convex optimization problem, and show that the optimum power-control is indeed binary under such approximations.  In \cite{Oh06}, the authors proved the optimality of an {\em almost} binary power-control strategy, up to one exceptional transmission power level in the continuum between $0$ and the peak power level, maximizing the total uplink communication rate.

The results reported in \cite{Gesbert08} as well as in other works raise the further question: When is ``binary'' power-control exactly optimal? It has been shown in very recent work \cite{Hanly10} that binary power-control is optimal when there is total symmetry amongst the links, {\em i.e.,} all direct link gains have one particular value, and all the cross-link gains have another particular value (possibly the same value as the direct link gain, but not necessarily).  One interesting feature of the result is that it is {\it as if} the sum-rate function of the powers were either Schur-convex, or Schur-concave (even though it is neither), leading to the observed result that either all links should be ``on'' or just one link should be ``on'' at the optimal solution.  A two-link Schur-convex/Schur-concave structure is observed and used, but it does not generalize to more than two links.

In the present paper, we study the sum-rate maximization problem for the classical multiple access channel, where all the links terminate in a common receiver node, but the link gains can be arbitrary.  In this setting, we show that the power-control problem can be solved quite easily via an underlying Schur-convex structure.  In contrast to the symmetric network of interfering links, it is no longer necessarily an all-or-one result: It is possible for the chosen set of links that are ``on'' to be larger than a singleton, but smaller than the set of all users, but it always consists of users with the best channels. On the other hand, we will observe from numerical results that the dominant modes, in terms of probability, correspond to the all-on or one-on solutions.

Majorization theory and Schur-convex/concave structures were also successfully utilized in some previous works, including \cite{Palomar06}, \cite{Viswanath99a}, \cite{Viswanath02} and \cite{Viswanath99b}, to answer important questions in communications theory.  This paper is another successful application of majorization theory to prove the optimality of binary power-control.

In \cite{Palomar06}, the authors focus on the transceiver design for point-to-point multiple-input-multiple-output (MIMO) communication systems.  By using extra degrees of freedoms provided by multiple transmitter and receiver antennas, and assuming either minimum mean-square error (MMSE) receiver or zero-forcing receiver, they show that the optimum linear precoder at the transmitter is the one diagonilazing the channels ({\it i.e.,} independent noise at all channels and no interference among them) when the cost function to be minimized is Schur-concave (or, the objective function to be maximized is Schur-convex).  Their results do not directly apply to the our problem since we consider the sum-rate maximization in the presence of interfering links in this paper.  In fact, we solve a special case of an open problem posed in \cite{Palomar06} in chapter 5 on the optimum design of transceivers for the MIMO multiple-access channel.

In \cite{Viswanath99a}, the authors focus on the design of capacity achieving spreading code sequences for the CDMA multiple-access channel without fading.  They allow multi-user detection for joint processing of users.  Even though the performance figure of merit we are interested in this paper is also related to the information capacity, our problem set-up is different than the set-up in \cite{Viswanath99a}.  In this paper, we look at the capacity achieving transmission power allocations, rather than the optimum spreading code sequence design, for Fading Gaussian channels in the presence of interfering links.  For example, our objective sum-rate function is Schur-convex whereas it is Schur-concave in \cite{Viswanath99a}.  In \cite{Viswanath02}, the same authors extend the analysis in \cite{Viswanath99a} to the case of colored noise.  In \cite{Viswanath99b}, they analyze the {\em user capacity}, which is defined as the maximum number of users that can be admitted to the system by allocating spreading code sequences and transmission powers optimally without violating minimum $\SINR$ requirements, of CDMA systems.  In this work, we focus on achieveable sum-rates rather than on user capacity.

Our results are different from the corresponding classic results in \cite{Knopp95}. In \cite{Knopp95}, the maximum Shannon-theoretic sum-rate is considered, whereas in the present paper, we treat interference as pure Gaussian noise. Although our assumption simplifies the receiver, it complicates the power optimization problem. We note that the capacity region of the Gaussian multiple-access channel is well understood, and it is known that all points of the boundary of the rate region can be achieved by successive decoding \cite{RU96}.  The optimal power-control for the Fading Gaussian multiple-access channel with channel state information at the transmitters is also well understood \cite{HanlyTse98a}.  In the present paper, we arrive at the problem from a different angle, where our interest is in understanding the structure of power-control problems in which interference is treated as Gaussian noise (very relevant for general interference networks), which excludes successive decoding or other multi-user decoding techniques.

From a practical perspective, treating interference as Gaussian noise is the approach taken in most existing systems, including cellular systems.  Note that the uplink of a cell is indeed a multiple-access channel.  Successive decoding is more complex to implement, and suffers from error propagation, which is mainly a problem if channels cannot be estimated very reliably.  We note that Qualcomm has recently produced a chip for successive decoding \cite{SZZ08}, so we cannot be sure that successive decoding will not be used in practice.  Indeed, we believe it will be.  In the present paper, we provide a comparison between the performance of the optimum binary power-control scheme with that of successive decoding, under various assumptions about the efficiency of the cancellation process. We expect that, in practice, successive decoding will be combined with user scheduling, to reduce the potential for error propagation, and the present paper provides insight into the problem of combined power-control and user scheduling, as will be shown.

\section{Network Model, Majorization and Nomenclature}
In this section, we will introduce the network model and some basic concepts from the theory of majorization.
\subsection{Network Model}
We focus on the uplink communication scenario where $n$ mobile users communicate with a single base station.  At time-slot $t$, the received signal at the base station is given by the baseband discrete-time Gaussian multiple-access channel as
\begin{eqnarray}
Y(t) = \sum_{i=1}^n \sqrt{h_i(t)} X_i(t) + W(t), \nonumber
\end{eqnarray}
where $X_i(t)$ and $h_i(t)$ are the transmitted signal and the channel fading coefficient of the $i^{\rm th}$ user, respectively, and $W(t)$ is white Gaussian noise with variance $\sigma^2$ at the base station.  We assume that $W(t)$ represents the cumulative effect of the thermal noise and other-cell interference at the base station.  Without loss of generality, we assume that all users are subject to the same peak transmission power constraint of $P$, {\em i.e.,} $\ES{|X_i(t)|^2} \leq P$ for all $t$.\footnote{If the users in the original rate maximization problem have different peak transmission power constraints given by the peak power vector $\vec{P} = \paren{P_1, \cdots, P_n}^\top$, then solving the modified optimization problem having the uniform peak power constraint $P$  and the fading processes that are scaled versions of the ones in the original problem by a factor of $\frac{P_i}{P}$, for all $i \in \brparen{1, \cdots, n}$, will be enough to find the optimal transmission power allocation for the original problem.}  We call a power allocation vector (at time-slot $t$) $\vec{P} = \paren{P_1, \cdots, P_n}^\top$ {\em binary} if $P_i$ is either $P$ or $0$ for all $i$.\footnote{If there is a minimum transmission power $P_{\min}$ requirement to maintain some level of control traffic in the network, then $\vec{P}$ is defined to be binary if $P_i$ is either $P$ or $P_{\min}$ for all $i$.} The signal-to-noise-ratio ($\SNR$) of the communication system under consideration is defined to be the ratio $\rho = \frac{P}{\sigma^2}$.

In Section \ref{Section: Optimal Binary Power Control}, we will solve the optimum power allocation problem for time-invariant (slow fading) channels characterized by a fixed channel vector $\vec{h}$, {\em i.e.,} $h_i(t) = h_i$ for all $t$.  Extensions to time-varying (fast fading) channels are straightforward.

\subsection{Majorization and Nomenclature}
$\R^m$ and $\Rp^m$ represent the set of $m$ dimensional column vectors with real and real non-negative coordinates, respectively.  For a vector $\vec{x}$ in $\R^m$, we denote its ordered coordinates by $\imax{x}{1} \geq \cdots \geq \imax{x}{m}$, and $\diag{\vec{x}}$ represents the diagonal matrix with entries of $\vec{x}$ at the diagonal.  When we write $\1$ (in boldface), we mean the vector of ones.  For $\vec{x}$ and $\vec{y}$ in $\R^m$, we say $\vec{x}$ {\em majorizes} $\vec{y}$ and write it as $\vec{x} \major \vec{y}$ if we have $\sum_{i=1}^k \imax{x}{i} \geq \sum_{i=1}^k \imax{y}{i}$ when $k = 1, \cdots, m-1$, and $\sum_{i=1}^m \imax{x}{i} = \sum_{i=1}^m \imax{y}{i}$.

A function $g: \R^m \mapsto \R$ is said to be {\em Schur-convex} if $\vec{x} \major \vec{y}$ implies $g\paren{\vec{x}} \geq g\paren{\vec{y}}$;  $g$ is said to be {\em strictly Schur-convex} if $g$ is Schur-convex, and $\vec{x} \major \vec{y}$ implies $g\paren{\vec{x}} > g\paren{\vec{y}}$ for all $\vec{x}$ and $\vec{y}$ which are not a permutation of each other.  $g$ is Schur-concave if $-g$ is Schur-convex.  Intuitively, a Schur-convex function increases when the dispersion among the components of its argument increases.

Schur-convex/concave functions frequently arise in mathematical analysis and engineering applications, {\em e.g.,} \cite{Palomar06}, \cite{Viswanath99a}, \cite{Viswanath02}, \cite{Viswanath99b} and \cite{Arnold07}.  For example, every function that is convex and symmetric is also a Schur-convex function.  Another important example of a Schur-convex function is a separable-convex function.  A function $g: \mathcal{I}^m \mapsto \R$, where $\mathcal{I} \subseteq \R$ is an interval, is said to be a {\em separable-convex function} if $g$ is of the form $g(\vec{x}) = \sum_{i=1}^m f\paren{x_i}$, where $f$ is a convex function on $\mathcal{I}$.  Then, any separable-convex function is also a Schur-convex function. (See \cite{Olkin79} or \cite{Arnold87}.)

\section{Main results} \label{Section:main results}

\subsection{Optimality of Binary Power-control} \label{Section: Optimal Binary Power Control}
In this section, we will prove the optimality of binary power-control for single cell communication systems without successive decoding at the base station.  
We begin by assuming that the channel is time-invariant and characterized by a fixed channel vector $\vec{h} \in \Rp^n$ given at time $0$.  The vector $\vec{h}$ can be generated according to a probability distribution, but once it is generated, it is fixed and known by the base station. For this case, we drop the time index, and write the sum-rate per slot as
\begin{eqnarray}
R_{\vec{h}}(\vec{P}) = \frac{1}{2} \sum_{i=1}^n \log\paren{1+\frac{h_i P_i}{\sigma^2+\sum_{j=1}^n h_j P_j \I{j \neq i}}}, \label{Eqn: Time-invariant sum-rate}
\end{eqnarray}
where $\vec{P} = \paren{P_1, \cdots, P_n}^\top$ is the vector of transmission powers.  The base of the logarithm function in
(\ref{Eqn: Time-invariant sum-rate}) is equal to the natural number $\e{}$, and therefore communication rates in this paper are measured in terms of nats per time-slot.

The sum-rate in (\ref{Eqn: Time-invariant sum-rate}) can be achieved using Gaussian input distributions and random coding arguments, and this is the focus of the present paper.  In general, these rates are not optimal, and higher rates in the multi-user capacity region are known to be achievable \cite{CT06}.  In fact, there is nothing inherently suboptimal about using Gaussian codebooks: The suboptimality of \eqref{Eqn: Time-invariant sum-rate} comes from a failure to exploit the information content in the interference, which can be removed via cancellation.  Nevertheless, we will treat the interference as Gaussian noise in the present paper, and in this context the relevant achievable rates are given in \eqref{Eqn:
Time-invariant sum-rate}.

We are interested in solving the following {\em non-convex} optimization problem.
\begin{eqnarray}
\begin{array}{ll}
\mbox{maximize} & R_{\vec{h}}(\vec{P}) \\
\mbox{subject to} & \vec{P} \preceq P \1
\end{array}. \label{Eqn: Problem 1}
\end{eqnarray}
Even though $R_{\vec{h}}(\vec{P})$ is a non-convex function of transmission powers, it is a strictly Schur-convex function of {\it received} powers at the base station, which will enable us to obtain the solutions for the non-convex optimization problem in (\ref{Eqn: Problem 1}).
\begin{lemma} \label{Lemma: Strict Schur-convexity}
Let $\dom = \bigotimes_{i=1}^n \sqparen{0, h_i P}$, $\vec{x} = \diag{\vec{P}} \cdot \vec{h}$ ({\em i.e.,} $\vec{x}$ changes as $\vec{P}$ changes), and write $R_{\vec{h}}(\vec{x})$ as a function of $\vec{x} = \paren{x_1, \cdots, x_n}^\top$ as
\begin{eqnarray}
R_{\vec{h}}(\vec{x}) = \frac{1}{2}\sum_{i=1}^n \log\paren{1+\frac{x_i}{\sigma^2+\sum_{j=1}^n x_j \I{j \neq i}}}. \label{Eqn: Instantaneous Rate}
\end{eqnarray}
Then, $R_{\vec{h}}(\vec{x})$ is a strictly Schur-convex function of $\vec{x}$ on $\dom$.
\end{lemma}
\proof Fix $B \geq 0$, and define $\dom_B = \brparen{\vec{x} \in \R^n: \vec{x} \in \dom \mbox{ and } \sum_{i=1}^n x_i = B}$.  On $\dom_B \neq \emptyset$, we can write $\insrate{\vec{h}}{\vec{x}}$ as
\begin{eqnarray}
\insrate{\vec{h}}{\vec{x}} = \frac12 \sum_{i=1}^n \log\paren{\frac{\sigma^2 + B}{\sigma^2 + B - x_i}}. \nonumber
\end{eqnarray}

We define $g\paren{\vec{y}} = \frac12 \sum_{i=1}^n \log\paren{\frac{\sigma^2 + B}{\sigma^2 + B - y_i}}$ on $\sqparen{0, B}^n$.  Note that $g\paren{\vec{y}}$ is a separable-convex function on $\sqparen{0, B}^n$ since $\log\paren{\frac{\sigma^2 + B}{\sigma^2 + B - y}}$ is a strictly convex function on $\sqparen{0, B}$. Thus, we conclude that $g\paren{\vec{y}}$ is strictly Schur-convex on $\sqparen{0, B}^n$.  Since $R_{\vec{h}} \equiv g$ on $\dom_B$, we also conclude that $R_{\vec{h}}$ is a strictly Schur-convex function on $\dom_B$ for any $B \geq 0$ such that $\dom_B \neq \emptyset$.  Since $\dom = \bigcup_{B \geq 0}\dom_B$, this last observation further implies that $R_{\vec{h}}$ is a strictly Schur-convex function on $\dom$. \qed

Note that $\vec{x}$ is in $\dom$ if and only if $\vec{P} \preceq P \1$.  Therefore, maximizing $R_{\vec{h}}(\vec{x})$ on $\dom$ is equivalent to solving the optimization problem in (\ref{Eqn: Problem 1}).  This observation together with the Schur-convexity of $R_{\vec{h}}$ will be the key for characterizing the optimum power allocation vectors.

The following are two simple facts about an optimum power allocation vector $\vec{P}^*$ solving (\ref{Eqn: Problem 1}).  At $\vec{P}^*$, there must exist at least one user transmitting with positive power, and if there is only one user transmitting with positive power, this user must transmit with full power.  It also directly follows from the Schur-convexity of $R_{\vec{h}}$ that if there are more than one users transmitting with positive power, one of them must transmit with full power.\footnote{This can also bee seen by using simple scaling arguments \cite{Gesbert08}.}  Otherwise, we can majorize the received power vector $\vec{x} = \diag{\vec{P}^*} \cdot \vec{h}$, and obtain a strictly better sum-rate by re-adjusting transmission powers without violating the transmission power constraint.  The next theorem establishes the binary nature of $\vec{P}^*$ and its structural properties.
\begin{theorem} \label{Thm: Optimum Power Control 1}
Any $\vec{P}^*$ solving the problem (\ref{Eqn: Problem 1}) is a {\em binary} power allocation vector at which the users transmitting with full power correspond to the ones having the best channel gains.
\end{theorem}
\IEEEproof:  See Appendix \ref{App: Proof of Optimum PC}. \qed

We now address the issue of uniqueness. Let $\pc\paren{\vec{h}} = \paren{P_1(\vec{h}), \cdots, P_n(\vec{h})}^\top$ be {\it any} optimal binary power allocation. Note that this definition extends the model to allow fading, and we can consider $\pc\paren{\vec{h}}$ as providing a power control policy, adaptive to changing channel conditions. Then the following theorem provides uniqueness.

\begin{theorem} \label{Thm: Optimum Power Control 2}
Any optimal power-control policy $\pc^*(\vec{h})$ assigns the channel to the best users for almost all fading states.  If the stationary distribution of the fading process is absolutely continuous, then $\pc^*(\vec{h})$ is unique up to a set of measure zero.
\end{theorem}
\IEEEproof See See Appendix \ref{App: Proof of Uniqueness}. \qed

We note that the set of optimum power allocation vectors solving (\ref{Eqn: Problem 1}) is not necessarily a singleton.  However, Theorem $\ref{Thm: Optimum Power Control 2}$ establishes uniqueness if the channel state vector is generated by an absolutely continuous distribution, which is a valid assumption for most practical systems.  Therefore, when we refer to an optimum power allocation vector or power-control policy for the rest of the paper, we will use $\pc^*$-notation without any ambiguity.



Finally, it is important to consider what the constraint in \eqref{Eqn: Problem 1} means in the case of a fading channel. We can interpret this constraint as a {\it peak} power constraint. If $P$ were an average power constraint on the powers modulating Gaussian codebooks \cite{HanlyTse98a}, then we would replace the constraint that $\displaystyle \pc(\vec{h}) \preceq P \1 \mbox{ for all } \vec{h} \in \Rp^n$ with the less onerous constraint that $\displaystyle {\mathbb E}[\pc(\vec{h})] \preceq P \1$. The reason for interest in peak power constraints is that in practice it is necessary to operate within the linear range of a power amplifier, and this may preclude bursts of power that may be required if only the average power is constrained.

\subsection{Polynomial-time Algorithm for Finding $\pc^*$}

In this section, we provide a polynomial-time algorithm for finding the optimum power allocation vector $\pc^*(\vec{h})$ for a given channel state vector $\vec{h}$.  One of the consequences of the structure of the optimum power-control policy established above is that it is piecewise constant: There exists a partition of the fading state space into $2^n-1$ regions upon each of which the optimum power-control policy is constant:
\begin{eqnarray}
\pc^*(\vec{h}) = \sum_{\mathcal{S} \subseteq \brparen{1, \cdots, n}, \atop \mathcal{S} \neq \emptyset} \vec{P}_{\mathcal{S}} \I{\vec{h} \in \mathcal{D}_{\mathcal{S}}}, \nonumber
\end{eqnarray}
where $\vec{P}_{\mathcal{S}} = \paren{P_1, \cdots, P_n}^\top$ is a transmission power profile such that $P_i = P \I{i \in \mathcal{S}}$, and the $\mathcal{D}_{\mathcal{S}}$ is the region on which only the users in $\mathcal{S}$ transmit with full power, and the rest are not scheduled for transmission.  Even though it is possible to give exact characterizations of these optimum power-control regions when there are only a few users (e.g., see the two-user example in Section \ref{Section: TDMA}), it becomes prohibitively complex to determine them when there are many users.

On the other hand, the structure of the optimum binary power allocation established above allows us to construct a simple, polynomial-time algorithm to compute the optimum power profile for any realized fading state and any number of users in the cell, which can be hard-coded into a scheduler circuit, without the need for any explicit characterization of the optimum power-control regions.  The suggested algorithm takes a fading state $\vec{h}$ as an input, computes the sum-rates $R_k(\vec{h})$ at which the best $k$, $1 \leq k \leq n$, users transmit with full power, and returns the optimum sum-rate maximizing transmission power profile at which only the best $k^*$ users are scheduled for transmission with full power. The
pseudocode for this simple polynomial-time algorithm is shown below.

\begin{algorithm}
\caption{Algorithm for computing optimum power allocation}
\label{Algorithm: Power Allocation}
\mbox{{\bf Input:} Fading state $\vec{h} \in \R^n$} \\
\mbox{{\bf Output:} Max. sum-rate $R_{\vec{h}}(\pc^*)$ and opt. power profile $\pc^*  \in \Rp^n$}
\begin{algorithmic}
\STATE {\bf Initialization:} $R_1\paren{\vec{h}} := \frac12 \log\paren{1+\rho \imax{h}{1}}$, $k^* := 1$, $R_{\vec{h}}(\pc^*) := R_1\paren{\vec{h}}$
\FOR {$k=2$ to $n$}
    \STATE $R_k(\vec{h}) = \frac12 \sum_{i=1}^k \log\paren{1+\frac{\imax{h}{i}}{\rho^{-1} + \sum_{j=1}^k \imax{h}{k} \I{j \neq i}}}$
    \IF {$R_k(\vec{h}) > R_{\vec{h}}(\pc^*)$}
        \STATE $R_{\vec{h}}(\pc^*) = R_k(\vec{h})$, $k^* = k$
    \ENDIF
\ENDFOR
\RETURN {\bf (i)} Max. sum rate:  $R_{\vec{h}}(\pc^*)$. {\bf (ii)} $\pc^*$: allocate TX power $P$ to the best $k^*$ users, and zero to the rest.
\end{algorithmic}
\end{algorithm}

\section{When is TDMA optimal?} \label{Section: TDMA}
In this section, we will establish the conditions under which the channel-state aware TDMA policy, in which the channel is allocated to the best user, is optimal for maximizing sum-rate in single cell wireless communication systems.  Optimality of this TDMA policy was established (under symmetric fading distributions) in previous works such as \cite{HanlyTse98a} and \cite{Knopp95} when even successive decoding for interference cancellation is allowed, and users are subject to an {\em average} power constraint.  On the other hand, as Theorems \ref{Thm: Optimum Power Control 1} and \ref{Thm: Optimum
Power Control 2} suggest, this TDMA policy is not always optimal in the communication scenario considered in this paper where successive decoding is not allowed, and users are subject to peak power constraints. The following two-user example further illustrates this point quantitively.
\begin{example} \label{Example: Two user}
When there are two users in the system, the sum-rate maximizing power allocation $\pc^*\paren{\vec{h}}$ is either $\paren{P, 0}^\top$, $\paren{0, P}^\top$, or $\paren{P, P}^\top$ for any given fading state $\vec{h} = \paren{h_1, h_2}^\top$ by Theorem~\ref{Thm: Optimum Power Control 2}.
Writing down the aggregate communication rate expressions for all three cases separately, and comparing them, one can derive the following conditions for the optimal power allocation for the two-user communication scenario:
\begin{eqnarray}
\pc^*\paren{\vec{h}}^\top = \left\{\begin{array}{cc}
\paren{P, 0}^\top & \mbox{ if } h_1 > \rho^{-1} \sqrt{1 + h_2 \rho} \mbox{ and } h_1 \geq h_2 \\
\paren{0, P}^\top & \mbox{ if } h_2 > \rho^{-1} \sqrt{1 + h_1 \rho} \mbox{ and } h_2 > h_1 \\
\paren{P, P}^\top & \mbox{ if } h_1 \leq \rho^{-1} \sqrt{1+h_2 \rho} \mbox{ and } h_2 \leq \rho^{-1} \sqrt{1 + h_1 \rho}
\end{array}. \right.
\end{eqnarray}

These three optimum power allocation regions are illustrated in Fig. \ref{Fig: two-user policy}.  For any fading state $\vec{h}$ lying inside the shaded region in Fig. \ref{Fig: two-user policy}, the TDMA policy becomes suboptimal, and the sum-rate is maximized by allocating the full transmission power to both users.  This situation occurs when both users experience similar and severe channel conditions, {\it i.e.,} $h_i \leq \rho^{-1} \frac{1+\sqrt{5}}{2}, i=1, 2$.  On the other hand, if the channel conditions experienced by users are relatively different from each other, or any of them is good enough, {\it i.e.,} $h_i >  \rho^{-1} \frac{1+\sqrt{5}}{2}$, then the TDMA policy maximizes the sum-rate.

\begin{figure}[tbp]
\psfrag{h1}{\small $h_1$} \psfrag{h2}{\small $h_2$} \psfrag{h}{\small $\vec{h}$}
\psfrag{n1}{\small $\rho^{-1}\frac{1+\sqrt{5}}{2}$} \psfrag{n2}{\small $\rho^{-1}$}
\psfrag{PP}{\small $\pc^*(\vec{h}) = (P, P)^\top$} \psfrag{0P}{\small $\pc^*(\vec{h}) = (0, P)^\top$}
\psfrag{P0}{\small $\pc^*(\vec{h}) = (P, 0)^\top$} \psfrag{l1}{\small $h_2 = \rho^{-1} \sqrt{1+h_1\rho}$}
\psfrag{l2}{\small \hspace{-0.75cm} $h_1 = \rho^{-1} \sqrt{1+h_2\rho}$} \psfrag{l3}{\small $h_2 = h_1$}
\begin{center}
\includegraphics[scale=0.7]{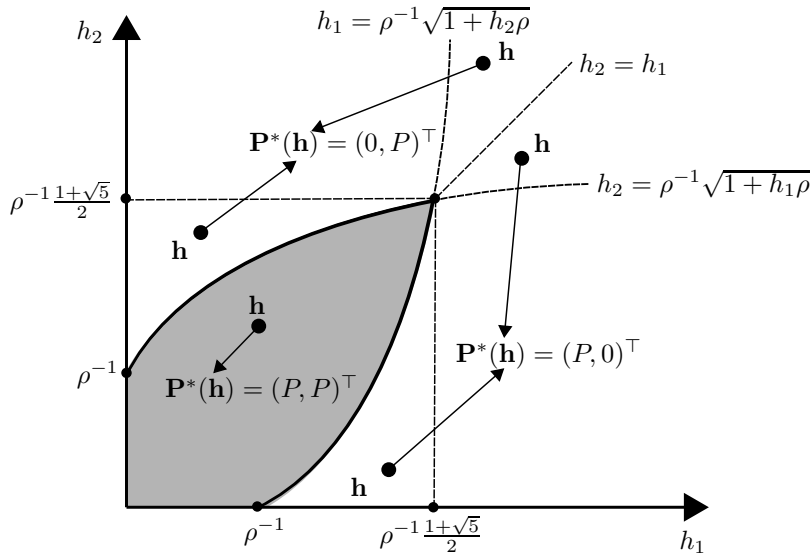}
\end{center}
\caption{Optimum power allocation regions for the two-user communication scenario.  For fading states lying in the shaded area, the TDMA policy is not optimal, and the sum-rate is maximized when both users transmit with full power.} \label{Fig: two-user policy}
\end{figure}

Note that the shaded region on which the TDMA policy is suboptimal shrinks to a point in the high $\SNR$ regime when $\rho$ grows to infinity.  Therefore, in the high $\SNR$ regime, we see one mode of communication with very high probability: Only the best user transmits with full power.  On the other hand, in the low $\SNR$ regime where $\rho$ goes to zero, the shaded region grows and covers the whole positive orthant in the $\R^2$-plane.  Therefore, in the low $\SNR$ regime, we again see only one mode of communication with very high probability:  All users transmit with full power.

When there are more than just two users, and for moderate $\SNR$ values, other modes of communication in which the best $k$, $1<k<n$, users transmit with full power can arise.  Roughly  speaking, the present discussion implies that the performance loss arising from the use of the TDMA policy for scheduling the best user critically depends on the relative strength of the peak transmission power with respect to the total noise power, including the background noise and other-cell interference, present in the system.  These observations will be the guiding principles for the proof of the optimality of the TDMA policy in the next theorem, and they will be further supported through numerical results in Section \ref{Section: Simulations}.

Figure \ref{Fig: two-user policy} also illustrates why $\pc^*$ is unique when the fading process has a continuous distribution.  When $\vec{h}$ lies on the boundary where any two of these three regions intersect, there are more than one power profile maximizing the sum-rate.  For example, all three power profiles $\paren{0, P}^\top, \paren{P, 0}^\top$ and $\paren{P, P}^\top$ perform equally well for sum-rate maximization at the point $\vec{h} = \paren{\rho^{-1}\frac{1+\sqrt{5}}{2}, \rho^{-1}\frac{1+\sqrt{5}}{2}}^\top$.  However, the probability of such a pathological case happening is zero, and $\pc^*$ can be almost surely uniquely determined if the joint stationary distribution of the fading process is absolutely continuous.
\end{example}

\begin{theorem} \label{Thm: TDMA Optimality}
For all $n \geq 1$, if $\imax{h}{1} \geq \paren{\e{} - 1} \rho^{-1}$ for a fading state $\vec{h}$, then the channel-state aware TDMA policy in which the channel is assigned to the user with the best channel state maximizes the sum-rate at this fading state.
\end{theorem}
\IEEEproof See Appendix \ref{App: Proof of TDMA Optimality}. \qed

\section{Numerical Results and Discussions} \label{Section: Simulations}

\subsection{Optimal modes: WB and TDMA}

In spite of the relative simplicity of Algorithm~\ref{Algorithm: Power Allocation}, we note that its worst case complexity is $\BO{n^2}$ when there are $n$ users, due to the ordering of the channel states of users and the summations involved. In this section, we examine the sum-rate performance of the heuristically derived scheme that simply takes the best of two choices: Either all users on at full power, which we call the wideband strategy (WB), or, exactly one user on at full power (the best user), which we call the TDMA strategy. To test out how well this suboptimal strategy works, we use the following simulation model.

We consider a circular cell centered at the base station and having radius $5$ [unit distance] (usually in kilometers).  We focus on low, moderate and high density networks, and vary the $\SNR$ parameter between $-30$dB and $30$dB to identify the performance of the power-controlled single cell communication systems for a broad spectrum of network parameters.  The users are uniformly distributed over the network domain with node density $\lambda$ [nodes per unit area].  The fading model includes both slow-fading, modeled by means of the bounded path-loss function $\frac{1}{1+x^\alpha}$ for $\alpha>2$ \cite{ICPW09}, and Rayleigh fast-fading, modeled by means of independent unit exponential random variables.\footnote{The same conclusions continue to hold for different cell sizes, different path-loss models including the unbounded path-loss model and generalized fading models including log-normal shadowing and other possible random factors.}  All simulations are performed in C over at least $10^4$ independent network realizations to obtain average aggregate communication rate figures.

\begin{figure}[tbp]
\begin{minipage}[t]{80mm}
\begin{center}
\includegraphics[scale=0.43]{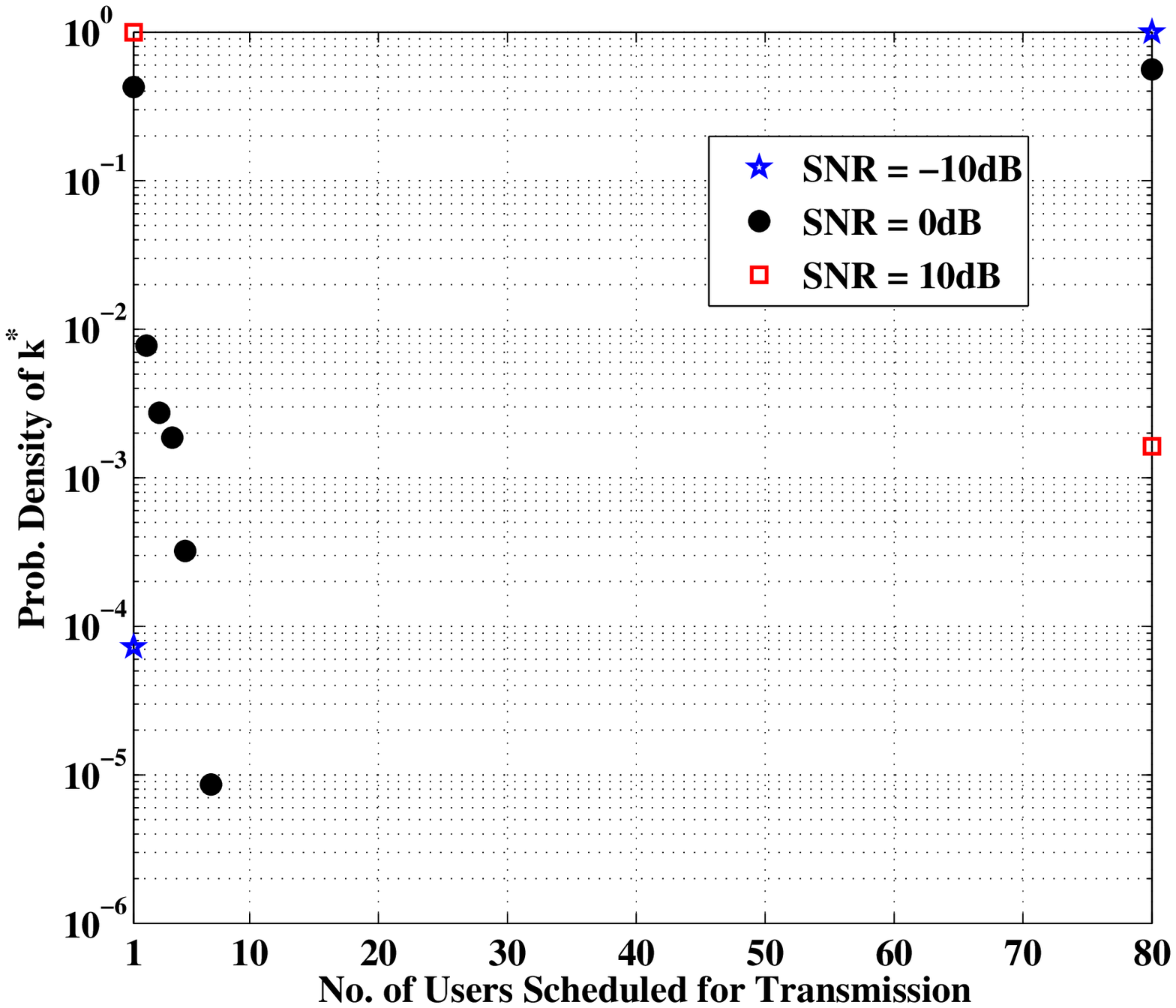}
\end{center}
\caption{Empirical probability density function of the optimum number of users scheduled for transmission. ($\lambda \approx 1$)} \label{Fig: Kopt1}
\end{minipage} \hspace{0mm}
\hspace{\fill}
\begin{minipage}[t]{80mm}
\begin{center}
\includegraphics[scale=0.43]{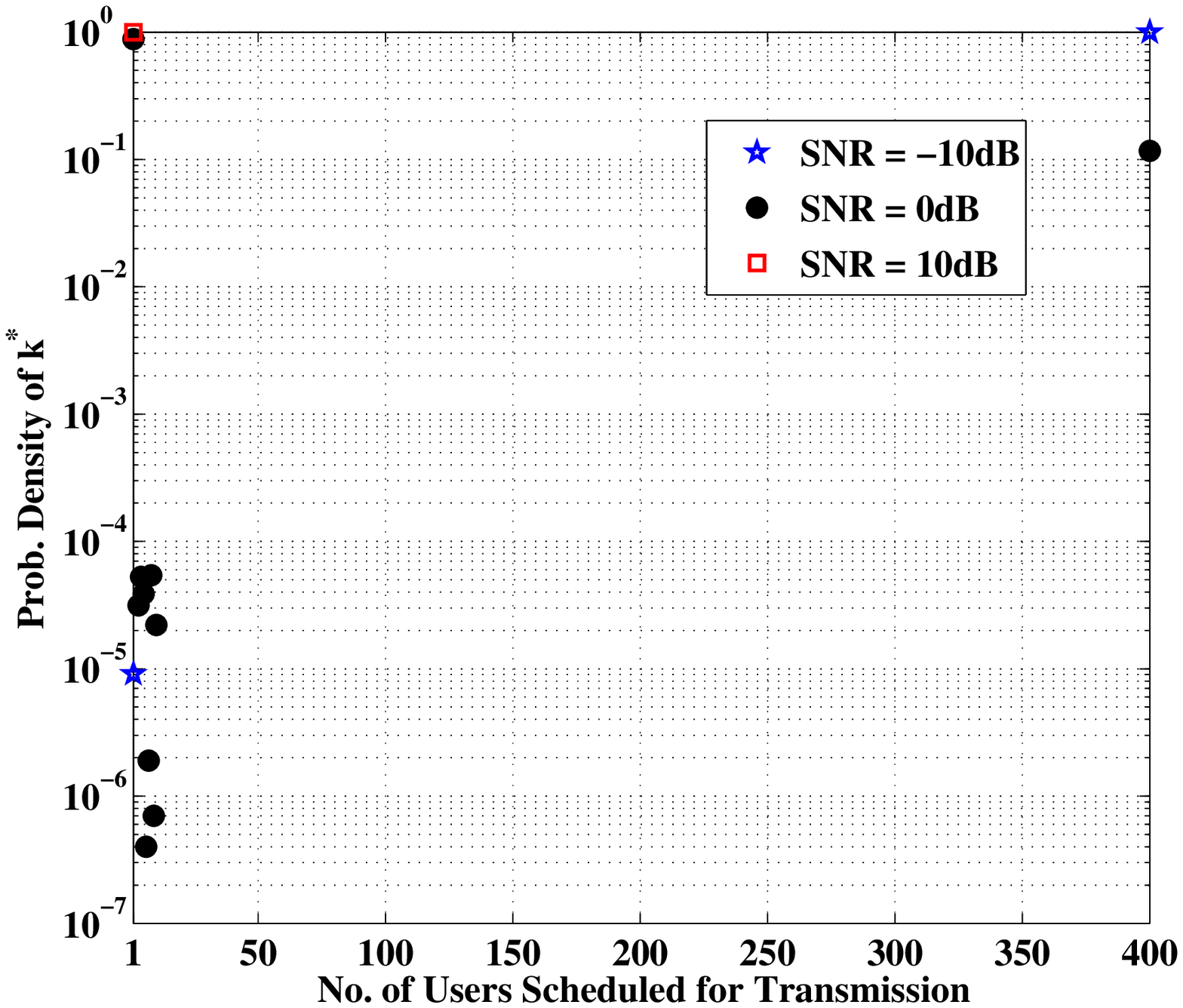}
\end{center}
\caption{Empirical probability density function of the optimum number of users scheduled for transmission. ($\lambda \approx 5$)}  \label{Fig: Kopt5}
\end{minipage}
\end{figure}

We begin by examining the empirical distribution of $k^*$, the number of users
scheduled in any fading state by Algorithm~\ref{Algorithm: Power Allocation} (the optimal algorithm). In Figs. \ref{Fig: Kopt1} and \ref{Fig: Kopt5}, we show
the empirical distribution obtained for $k^*$ over $10^7$
independent network realizations when $80$ ($\lambda \approx 1$) and
$400$ ($\lambda \approx 5$) users are uniformly distributed over the
network domain for $\SNR$ values $-10$dB, $0$dB and $10$dB.  Similar
conclusions continue to hold for different values of node density
and the $\SNR$ parameter.

In all cases, even though other modes of communication are quite possible, TDMA and WB modes predominantly arise.  The reason for such behavior is that when the channel state of the best user is good enough, we schedule just this user to maximize the communication rate; otherwise, the channels of the remaining users are also in deep fades, creating a domino effect and all users are scheduled together to maximize the communication rate. Similar observations were also made in \cite{Hanly10}, and proven to hold for the symmetric network of interfering links. Similarly, here, we can prove that scheduling a single user becomes certain as we scale up the node density. To see why this is so, consider first a model with a {\it fixed} number, $n$, of users, that we place uniformly at random in the cell. Since we have an {\it i.i.d.} model for the user locations, we can let $F(h)$ be the cumulative distribution function of the channel of a randomly selected user. Then the probability that {\it all} the users fail the condition of Theorem~\ref{Thm: TDMA Optimality} is $F^n\left(\paren{\e{} - 1} \rho^{-1}\right)$ which decays exponentially in $n$, irrespective of the SNR. Thus, for a large number of users we will almost certainly just schedule the best user, although the number of users required to observe this phenomena will be larger for lower SNR. It is a straightforward extension from this fixed $n$ model to the above numerical model, where the probability becomes ${\mathbb E}\left[F^N\left(\paren{\e{} - 1} \rho^{-1}\right)\right]$, where $N$ is the Poisson number of users with intensity $\lambda$, and one can show that this also decays exponentially in $\lambda$. This phenomena is illustrated in Figure~\ref{Fig: Kopt5} where only the best user is selected at SNR = 10 dB.

\begin{figure}[tbp]
\begin{minipage}[t]{80mm}
\begin{center}
\includegraphics[scale=0.43]{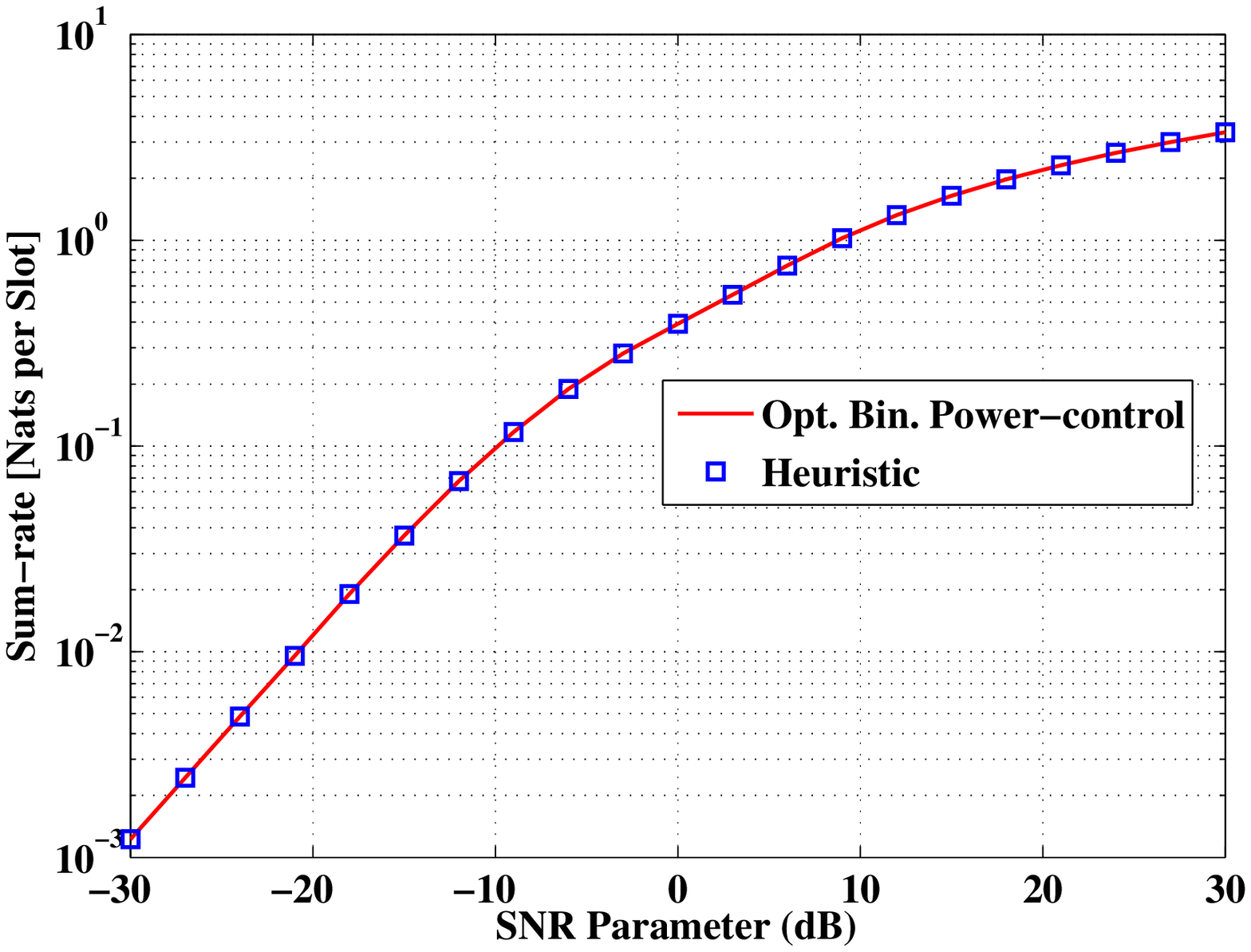}
\end{center}
\caption{Comparison of sum-rates achieved by the optimum binary power-control and the heuristic algorithm choosing either the TDMA mode or WB mode for transmission. ($\lambda = 0.5$)} \label{Fig: Heuristic05}
\end{minipage} \hspace{0mm}
\hspace{\fill}
\begin{minipage}[t]{80mm}
\begin{center}
\includegraphics[scale=0.43]{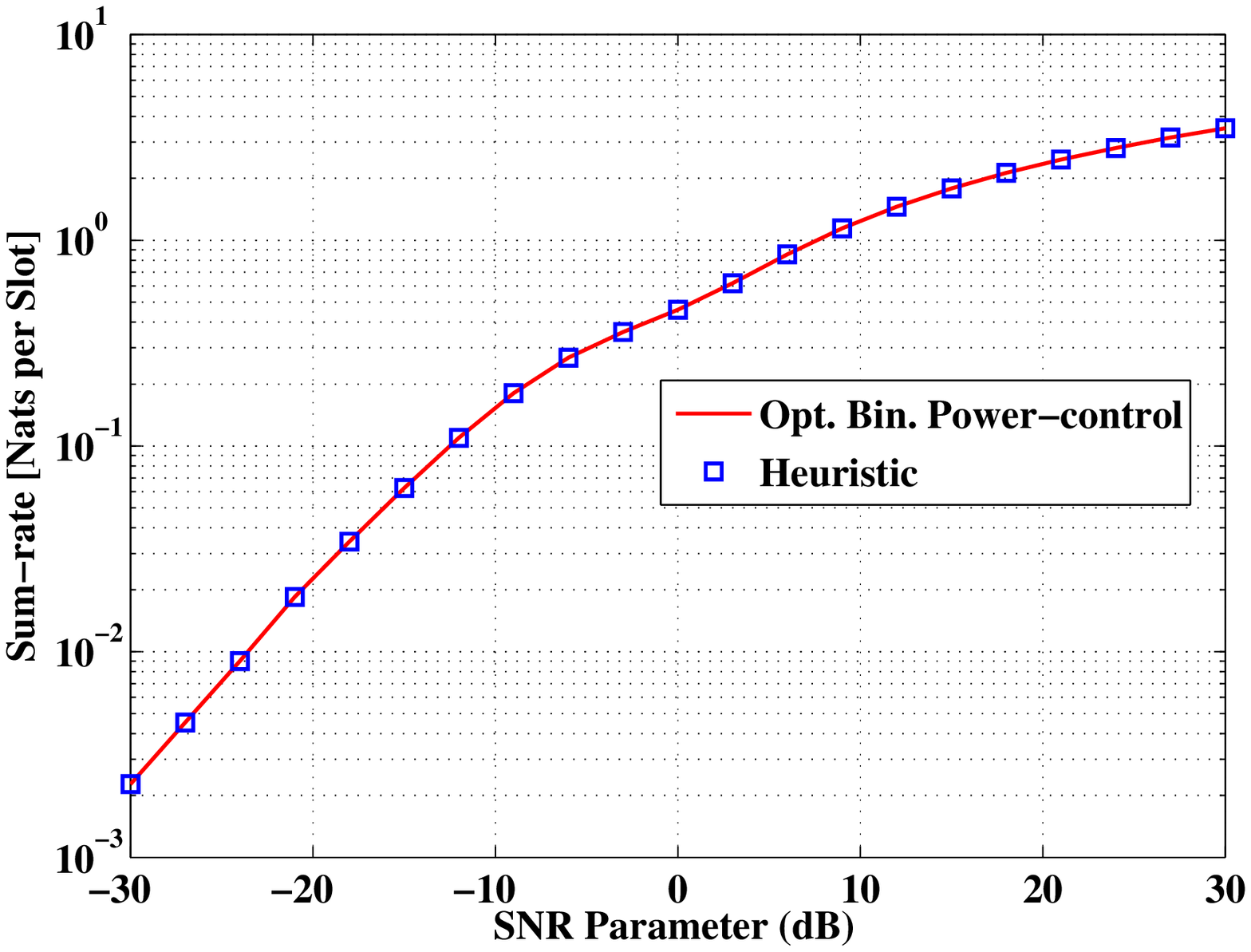}
\end{center}
\caption{Comparison of sum-rates achieved by the optimum binary power-control and the heuristic algorithm choosing either the TDMA mode or WB mode for transmission. ($\lambda = 1$)}  \label{Fig: Heuristic1}
\end{minipage}
\end{figure}

\begin{figure}[tbp]
\begin{minipage}[t]{80mm}
\begin{center}
\includegraphics[scale=0.43]{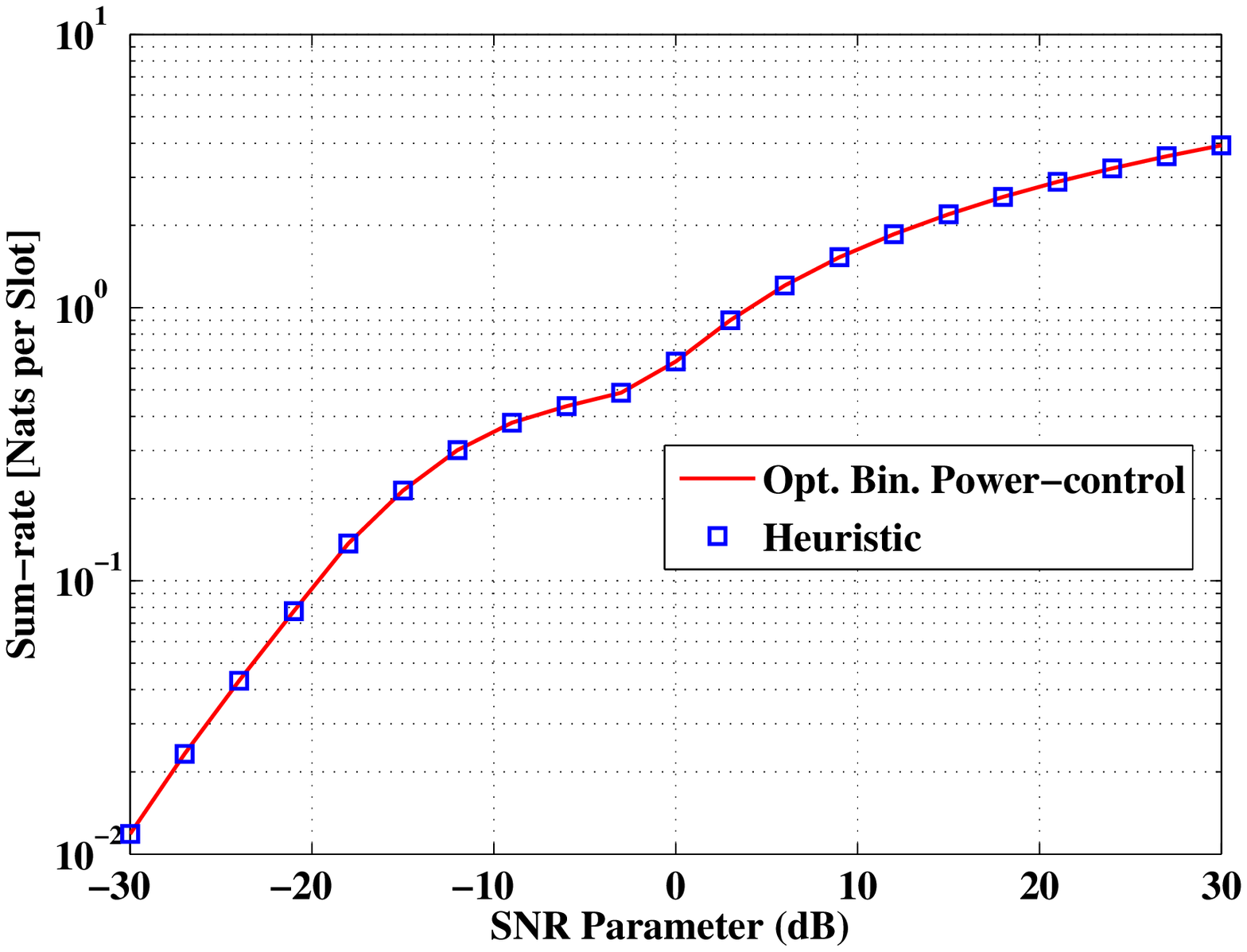}
\end{center}
\caption{Comparison of sum-rates achieved by the optimum binary power-control and the heuristic algorithm choosing either the TDMA mode or WB mode for transmission. ($\lambda = 5$)} \label{Fig: Heuristic5}
\end{minipage} \hspace{0mm}
\hspace{\fill}
\begin{minipage}[t]{80mm}
\begin{center}
\includegraphics[scale=0.43]{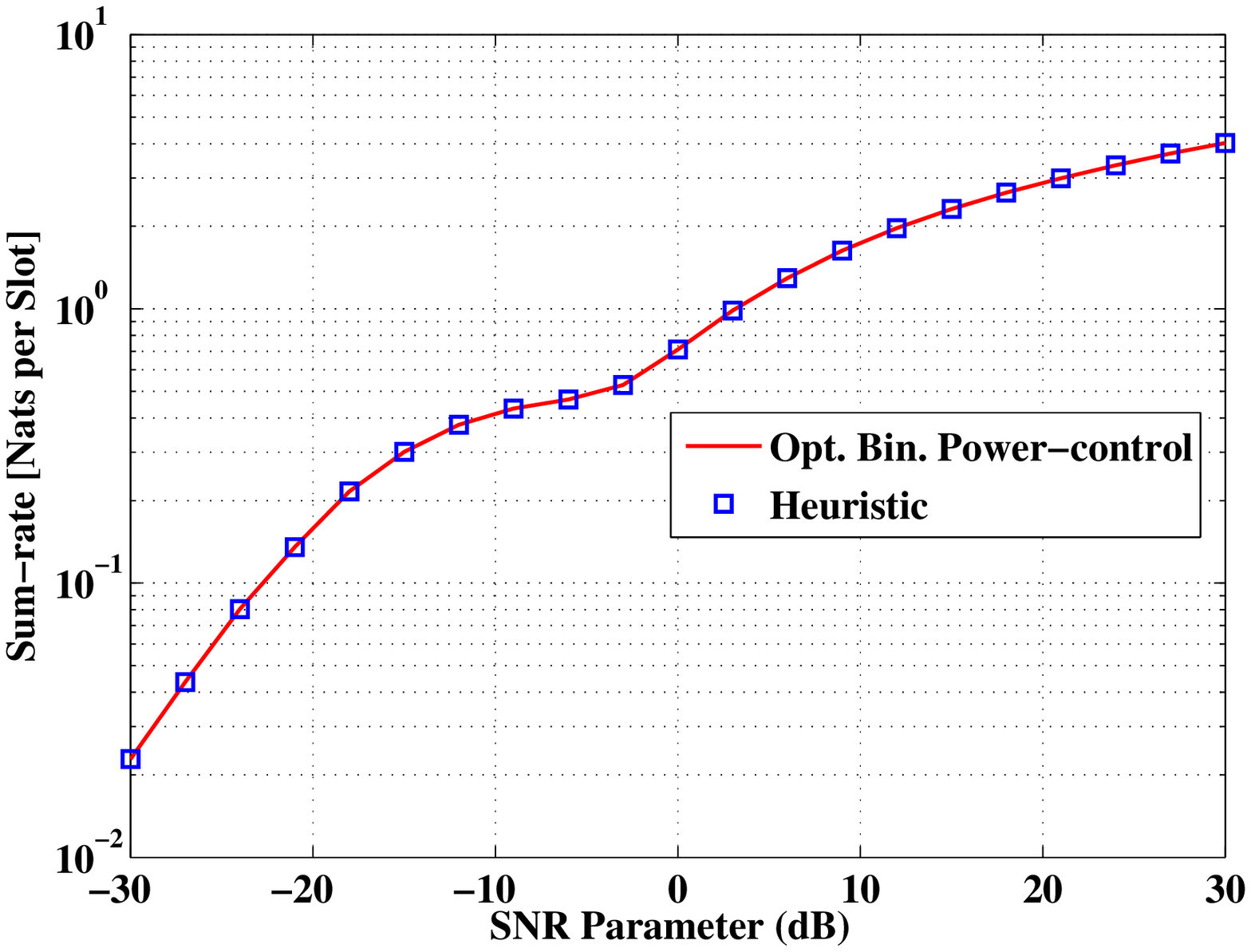}
\end{center}
\caption{Comparison of sum-rates achieved by the optimum binary power-control and the heuristic algorithm choosing either the TDMA mode or WB mode for transmission. ($\lambda = 10$)}  \label{Fig: Heuristic10}
\end{minipage}
\end{figure}

In Figs. \ref{Fig: Heuristic05}, \ref{Fig: Heuristic1}, \ref{Fig: Heuristic5} and \ref{Fig: Heuristic10}, we compare the sum-rates
achieved by the heuristic algorithm that simply chooses the best of the two extreme modes (WB or TDMA) with the rates achieved by the optimum binary power-control policy.  As illustrated in these figures, the performance achieved by the heuristic algorithm almost perfectly tracks the performance achieved by the optimum power-control, and therefore it can be implemented to maximize communication rates in single cell communication systems for all practical purposes without any noticeable performance degradation.  Especially, for systems with large numbers of users, the proposed heuristic algorithm will run an order of magnitude faster than Algorithm \ref{Algorithm: Power Allocation}. We also note that the
knee of the sum-rate curves (more apparent for high density networks) at which they become non-differentiable corresponds to a phase transition from the WB mode to the TDMA mode for scheduling users \cite{Hanly10}.

\subsection{Benefits from successive decoding}

In this section, we compare the aggregate communication rate achieved by the optimum binary power-control policy with the throughput capacity limits that can be achieved through successive decoding.  When the receiver is capable of successively decoding the received signals with cancellation efficiency $\beta \in [0, 1]$, which represents the amount of cancelled signal power, the throughput capacity can be given by
\begin{eqnarray}
C_{\rm SIC}(\beta) = \frac12 \EW_{\vec{h}}\sqparen{\sum_{i=1}^n\log\paren{1 + \frac{\imax{h}{i}}{\rho^{-1} + \sum_{j=1}^n \imax{h}{j} \I{j \neq i} - \beta \sum_{j=1}^{i-1} \imax{h}{j}}}}. \label{Eqn: Rate with SIC}
\end{eqnarray}
In (\ref{Eqn: Rate with SIC}), we used the usual decoding order in which the strongest users are decoded first and subtracted from the composite signal (see \cite{SZZ08}, \cite{Viterbi90} and \cite{HSPT06}).  Note that we obtain the classical throughput capacity equation $C_{\rm SIC}(1) = \frac12 \EW_{\vec{h}}\sqparen{\log\paren{1+\rho \sum_{i=1}^n h_i}}$ if the interference can be cancelled perfectly ($\beta=1$) \cite{HanlyTse98a}. Thus, there is no need for user scheduling when considering successive decoding under peak power constraints, and perfect channel state information at the base station. However, in practical implementations, $\beta$ is usually bounded away from one due to imperfect channel and signal estimations. In these cases, it may pay to do some user selection, but in the numerical results below, we assume that all users are scheduled for successive interference cancellation, as in \eqref{Eqn: Rate with SIC}.

\begin{figure}[tbp]
\begin{minipage}[t]{80mm}
\begin{center}
\includegraphics[scale=0.43]{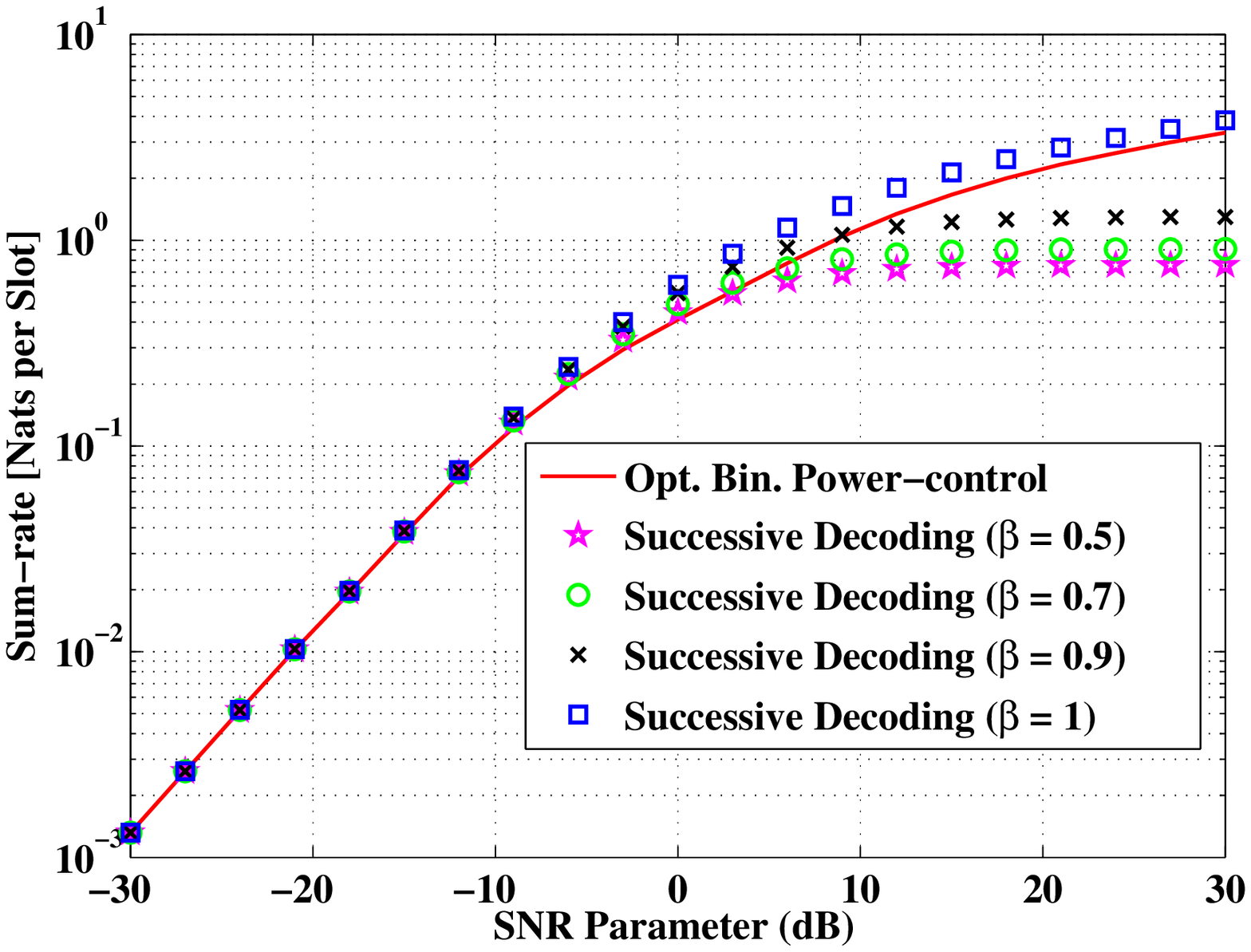}
\end{center}
\caption{Comparison of the sum-rate achieved by the optimum binary power-control and the throughput capacity limits achieved by successive decoding. ($\lambda = 0.5$)} \label{Fig: SIC05}
\end{minipage} \hspace{0mm}
\hspace{\fill}
\begin{minipage}[t]{80mm}
\begin{center}
\includegraphics[scale=0.43]{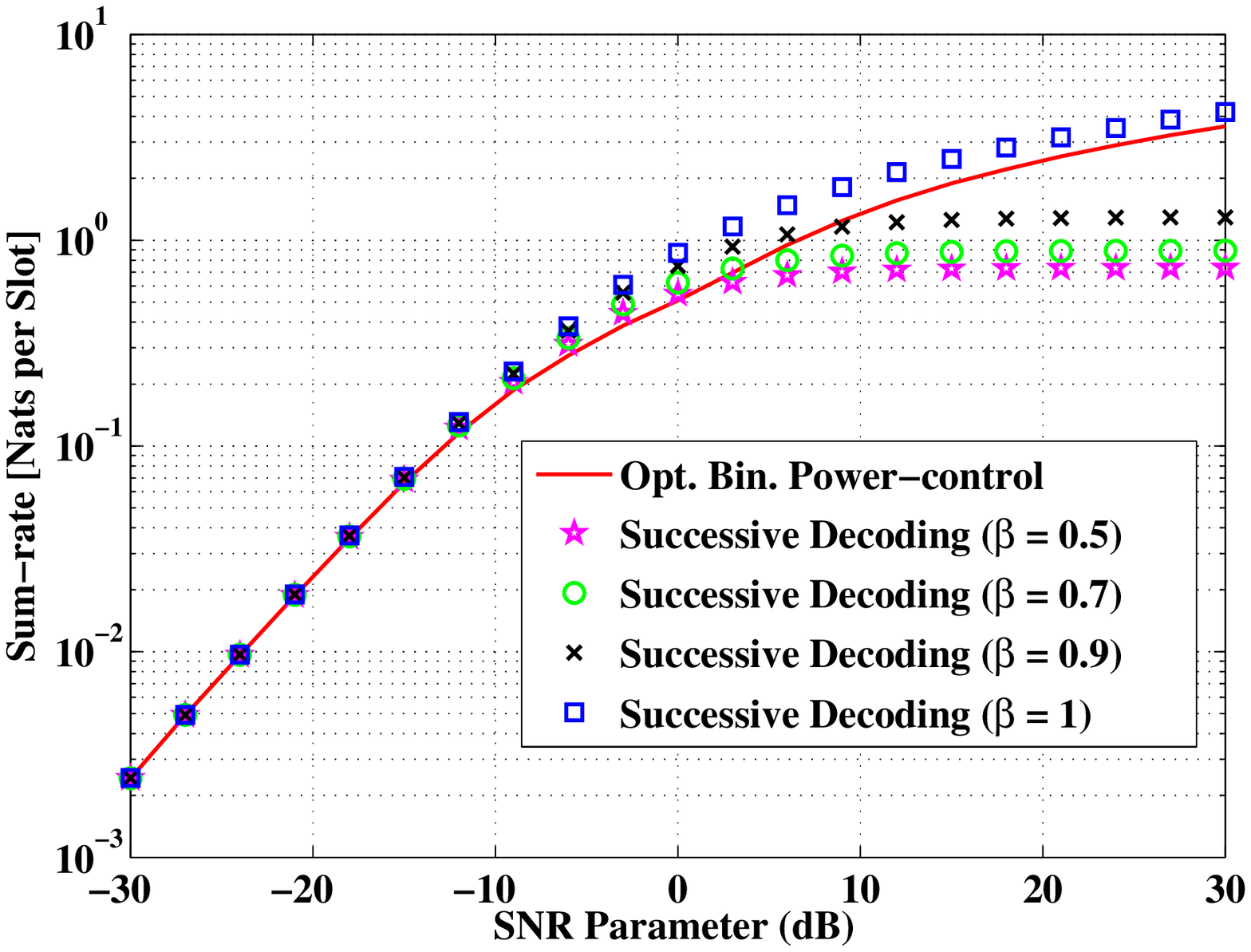}
\end{center}
\caption{Comparison of the sum-rate achieved by the optimum binary power-control and the throughput capacity limits achieved by successive decoding. ($\lambda = 1$)}  \label{Fig: SIC1}
\end{minipage}
\end{figure}

\begin{figure}[tbp]
\begin{minipage}[t]{80mm}
\begin{center}
\includegraphics[scale=0.43]{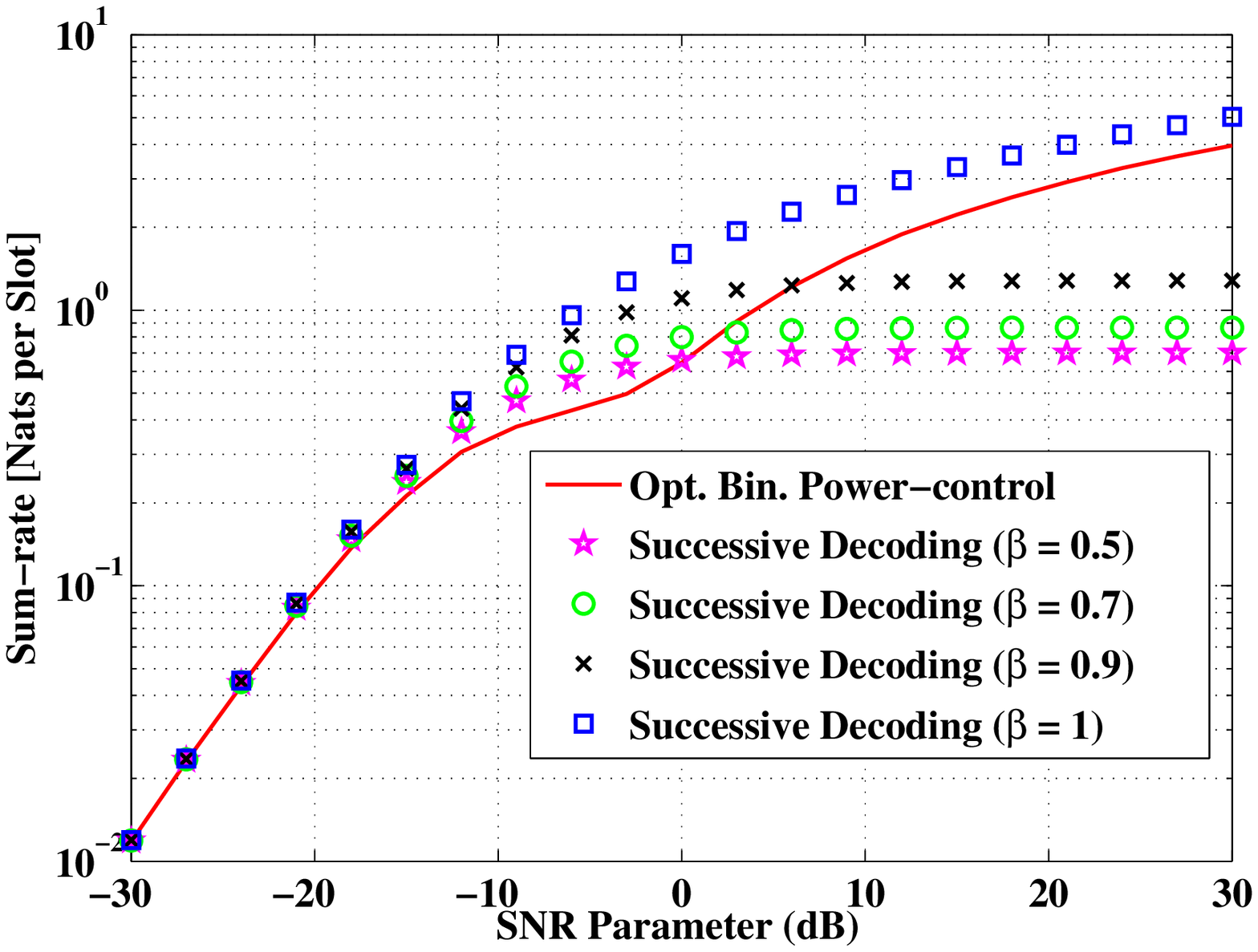}
\end{center}
\caption{Comparison of the sum-rate achieved by the optimum binary power-control and the throughput capacity limits achieved by successive decoding. ($\lambda = 5$)} \label{Fig: SIC5}
\end{minipage} \hspace{0mm}
\hspace{\fill}
\begin{minipage}[t]{80mm}
\begin{center}
\includegraphics[scale=0.43]{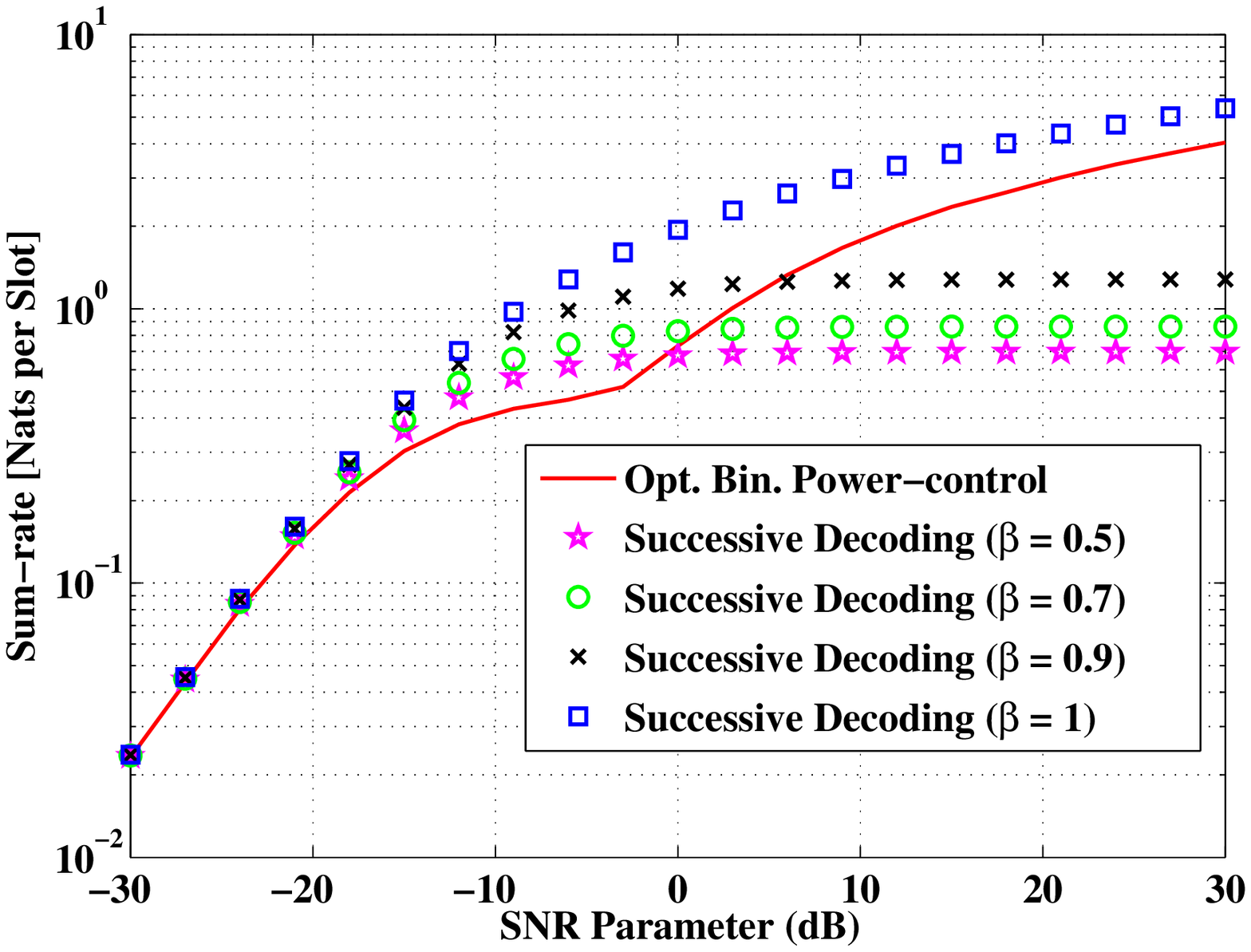}
\end{center}
\caption{Comparison of the sum-rate achieved by the optimum binary power-control and the throughput capacity limits achieved by successive decoding. ($\lambda = 10$)}  \label{Fig: SIC10}
\end{minipage}
\end{figure}

In Figs. \ref{Fig: SIC05}, \ref{Fig: SIC1}, \ref{Fig: SIC5} and
\ref{Fig: SIC10}, we depict the sum-rates achieved by the optimum
power-control policy and the throughput capacity limits achieved
through successive decoding.  As it must, the perfect successive
signal decoding capability increases the rates of communication that
can be achieved in single cell communication systems.  In
particular, for high density networks with moderate $\SNR$ values,
the performance increase achieved by the perfect successive decoding
can be as much as two times the average sum-rate achieved by the
optimal binary power-control treating all signals as noise.  On the
other hand, if the interference cancellation is not perfect and some
residual signal power remains after each cancellation step, the
sum-rate achieved by successive decoding saturates as $\SNR$
increases, and the optimum binary power-control can achieve higher
communication rates.  Therefore, practical successive interference
cancellations at the chip level (e.g., QUALCOMM CSM6850) require
near-perfect cancellation efficiency to harvest potential gains due
to complex successive decoding process.

In its favour, successive decoding does provide more fairness to users, as it enables all users to transmit and achieve sustainable data rates simultaneously. It is particularly well suited to the multiple cell context, as discussed in the conclusions section of \cite{HW93}, but we do not investigate that scenario in the present paper. Nor do we consider the impact of average power constraints, which may be very important in practice \cite{HanlyTse98a}.

\section{Conclusions}

This paper exploits the Schur-convexity property of the sum-rate function of received powers, to show that binary power-control is optimal for the multiple-access channel, when interference is treated as Gaussian noise, and there are peak power constraints on the users. If the fading distribution is absolutely continuous, then the optimum binary power-control policy is unique. We provide an algorithm to find the optimum power allocation, as a function of the channel state, that is polynomial in the number of users in the cell. However, we also present numerical results for a realistically dimensioned single cell system which suggest that there is essentially no loss in restricting attention to the best of two possible allocations in each channel state: (i) The best user transmits at peak power with other users switched off, as in channel-state aware TDMA, (ii) all users transmit simultaneously at peak power. This drastically reduces the complexity of the power allocation problem.   Finally, we compared all such schemes with successive decoding.

Our main conclusions regarding successive decoding are that as far as sum-rate maximization is concerned, successive decoding can gain up to about a factor of 2 over the optimal binary power-control scheme for the single cell model considered in the present paper, provided that the interference cancellation is perfect, and the SNR is moderate (not high or low). However, at high or low SNR, the gain is much less than that, and if the cancellation efficiency is less than 1 ({\it i.e.,} some small fraction of the interference remains) then the optimum binary power-control approach is superior, as it is not interference limited. It must be noted that this analysis pertains to only a single cell system, and to sum-rate maximization under peak power constraints. With multiple cells, and different objectives (such as maximization of logarithmic utilities) the conclusions are likely to be very different.

\appendices


\section{Proof of Theorem \ref{Thm: Optimum Power Control 1}} \label{App: Proof of Optimum PC}

We will first show that at $\vec{P}^*$, there cannot be two different users $i$ and $j$ with $0<P^*_i<P$ and $0<P^*_j<P$.  To obtain a contradiction, suppose there exist such two users.  Let $\vec{x} = \diag{\vec{P}^*} \cdot \vec{h}$, $x_i = h_i P^*_i$ and $x_j = h_j P_j^*$.  Since $\vec{P}^*$ is a solution for (\ref{Eqn: Problem 1}), we have $R_{\vec{h}}(\vec{x}) \geq R_{\vec{h}}(\vec{y})$ for all $\vec{y} \in \dom = \bigotimes_{i=1}^n \sqparen{0, h_i P}$.

Without loss of generality, assume $x_i \geq x_j$.  But now, we can re-adjust transmission power levels to achieve $0< y_i = x_i+\epsilon \leq h_i P$ and $0 \leq y_j = x_j - \epsilon < h_j P$ for some $\epsilon \geq 0$ small enough.  Then, the received power vector $\vec{y}$ formed as $y_i = x_i+\epsilon$, $y_j = x_j - \epsilon$ and $y_k = x_k$ for $k \neq i, j$, belongs to $\dom$ and majorizes $\vec{x}$.\footnote{$\vec{y} \major \vec{x}$ if and only if there exists a doubly-stochastic matrix $\vec{A}$ such that $\vec{x} = \vec{A} \vec{y}$.  We can construct $\vec{A}$ as follows.  For $k \neq i, j$, let $A_{k, l} = \I{l=k}, l \in \brparen{1, \cdots, n}$.  Let $A_{i,l} = a \I{l=i}, A_{i, l} = (1-a) \I{l = j}, A_{j, l} = (1-a) \I{l = i} \mbox{ and } A_{j, l} = a \I{l = j}, l \in \brparen{1, \cdots, n}$.  To find $a$, we solve for $\begin{pmatrix} a \\ 1 \end{pmatrix} = \begin{pmatrix} x_i - x_j +2\epsilon & x_j -\epsilon   \\ x_j - x_i - 2\epsilon & x_i + \epsilon \end{pmatrix}^{-1} \begin{pmatrix} x_i \\ x_j \end{pmatrix}$, which produces $a = \frac{x_i-x_j+\epsilon}{x_i-x_j+2\epsilon}$.}  By Lemma \ref{Lemma: Strict Schur-convexity}, $R_{\vec{h}}(\vec{y}) > R_{\vec{h}}(\vec{x})$, which produces a contradiction.  As a result, if $\vec{P}^*$ is a solution for (\ref{Eqn: Problem 1}), there can be at most one exceptional user with transmission power $c$ in $(0, P)$.  Others either transmit with full power, or do not transmit at all.

We will now show that this exceptional case does not happen.  Suppose $c \in (0, P)$.  Let $m$ be the index of the user with power $c$, and $\mathcal{S}$ be the subset of users transmitting with full power.  Let $H = \sum_{i \in \mathcal{S}} h_i$.  Then, $R_{\vec{h}}(\vec{x})$ on $\bigotimes_{i \in S} [0, h_iP] \bigotimes [0, h_m P]$ can be written as
\begin{eqnarray}
R_{\vec{h}}(\vec{x}) &=& \frac12 \sum_{i \in \mathcal{S}} \log\paren{1+\frac{x_i}{\sigma^2 + x_m + \sum_{j \in \mathcal{S}} x_j \I{j \neq i}}}  + \frac12 \log\paren{1+ \frac{x_m}{\sigma^2+\sum_{j \in S} x_j}} \nonumber \\
&=& \frac12 \sum_{i \in \mathcal{S}} \log\paren{1+\frac{h_i}{\rho^{-1} + H + \frac{c h_m}{P} - h_i}} + \frac12 \log\paren{1+\frac{\frac{c h_m}{P}}{\rho^{-1} + H}}. \nonumber
\end{eqnarray}
We define the following function on $[0, h_m]$.
\begin{eqnarray}
g(x) = \frac12 \sum_{i \in \mathcal{S}} \log\paren{1+ \frac{h_i}{\rho^{-1}+H-h_i+x}} + \frac12 \log\paren{1+\frac{x}{\rho^{-1} + H}}, \nonumber
\end{eqnarray}
whose derivative with respect to $x$ is
\begin{eqnarray}
g^\prime(x) = \frac12 \frac{1}{\rho^{-1} + H + x}\paren{1 - \sum_{i \in S} \frac{h_i}{\rho^{-1} + H - h_i + x}}. \nonumber
\end{eqnarray}
 $g$ has to be maximized at $x = \frac{c h_m}{P}$ because $\vec{P}^*$ solves (\ref{Eqn: Problem 1}).  Since $f(x) = 1 - \sum_{i \in S} \frac{h_i}{\rho^{-1} + H - h_i + x}$ is a strictly increasing function of $x$, we have $g^\prime(x)>0$ for $x > 0$ if $f(0) \geq 0$.  Thus, $g(h_m) > g\paren{\frac{c h_m}{P}}$, which is a contradiction.  If $f(h_m) \leq 0$, we have $g^\prime(x) < 0$ for $x < h_m$. Thus, $g(0)>g\paren{\frac{c h_m}{P}}$, which is a contradiction.  Similarly, if $f(h_m)>0$ and $f(0)<0$, we have $g\paren{\frac{c h_m}{P}} < \max\brparen{g(0), g(h_m)}$, which is another contradiction.  As a result, $c$ must be either zero or $P$, which proves that $\vec{P}^*$ is binary, and it strictly dominates any non-binary power allocation vector.

To see why the users with the best channel states transmit with full power, assume that $h_i>h_j$, $P_i^*=0$ and $P_j^*=P$.  We can achieve the same aggregate communication rate by setting the transmission power of the $i^{\rm th}$ user to $\frac{P h_j}{h_i} < P$ and that of the $j^{\rm th}$ user to zero.  However, such a transmission power allocation can be strictly dominated by a binary transmission power allocation as proven above.  Therefore,  users transmitting with full power correspond to the ones with the best channel states when transmission powers are allocated according to $\vec{P}^*$.

\section{Proof of Theorem \ref{Thm: Optimum Power Control 2}} \label{App: Proof of Uniqueness}
Binary structure of $\pc^*(\vec{h})$ directly follows from Theorem \ref{Thm: Optimum Power Control 1} and some measure theoretic arguments.  Therefore, we focus on the uniqueness of $\pc^*(\vec{h})$.  We define the sum-rate at a fading state $\vec{h}$ when the best $k$ users transmit with full power as
\begin{eqnarray}
R_k(\vec{h}) = \frac12 \sum_{i=1}^k \log\paren{1+\frac{\imax{h}{i}}{\rho^{-1} + \sum_{j=1}^k \imax{h}{j} \I{j\neq i}}}. \nonumber
\end{eqnarray}
We want to show that $\mathcal{S} = \brparen{\vec{h} \in \R^n: \exists k, m \mbox{ such that } k \neq m \mbox{ and }R_k(\vec{h}) = R_m(\vec{h}) }$ has probability zero with respect to the stationary distribution of the fading process.  To this end, it is enough to show that $\mathcal{S}$ has zero volume since the stationary fading distribution is absolutely continuous.  Suppose not.  Then, we can find $m>k$ such that $\mathcal{S}_{k, m} = \brparen{\vec{h} \in \R^n: R_k(\vec{h}) = R_m(\vec{h})}$ has positive volume.  First, let $m=k+1$.  This means that we can find a point $\vec{y} \in \mathcal{S}_{k, k+1}$ and a small $\R^{k+1}$-ball $\mathcal{B}\paren{\vec{y}, \epsilon} \subseteq \mathcal{S}_{k, k+1}$ centered around $\vec{y}$.  This implies that as a function of its largest $(k+1)^{\rm th}$ component (keeping other coordinates constant at $\imax{y}{i}$, $1\leq i \leq k$), $R_{k+1}(\vec{h})$ is constant over $\paren{\imax{y}{k+1} - \epsilon, \imax{y}{k+1} + \epsilon}$.  One can show that this cannot happen by taking the partial derivative of $R_{k+1}(\vec{h})$ with respect to $\imax{h}{k+1}$.

Similarly, if $m \geq k+2$, we can find a point $\vec{y} \in \mathcal{S}_{k, m}$ and a small $\R^m$-ball $\mathcal{B}\paren{\vec{y}, \epsilon} \subseteq \mathcal{S}_{k, m}$ centered around $\vec{y}$ such that $R_{m}(\vec{h})$ is constant over this ball as a function of its largest $(k+j)^{\rm th}$, $j=1, \cdots, m-k$, components.  However, by following the same steps in Lemma \ref{Lemma: Strict Schur-convexity}, it is not hard to show that $R_{m}(\vec{h})$ is a strictly Schur-convex function as a function of the largest $m$ elements of $\vec{h}$.  Therefore, $R_{m}(\vec{h})$ cannot be constant over $\mathcal{B}\paren{\vec{y}, \epsilon}$ as a function of its largest $(k+j)^{\rm th}$, $j=1, \cdots, m-k$, components since we can obtain a different $\vec{h}_1$ from a given $\vec{h}_2$, both in $\mathcal{B}\paren{\vec{y}, \epsilon}$, such that $\vec{h}_1 \major \vec{h}_2$ by only perturbing the largest $(k+j)^{\rm th}$, $j=1, \cdots, m-k$, components.


\section{Proof of Theorem \ref{Thm: TDMA Optimality}} \label{App: Proof of TDMA Optimality}
From a given fading state $\vec{h}$, we derive another fading state $\vec{g} = \1 \imax{h}{1}$ by making the channel conditions of all users the same and equal to $\imax{h}{1}$.  For these two fading states, we have 
\begin{eqnarray}
R_{\vec{g}}\paren{\pc^*} \geq R_{\vec{h}}\paren{\pc^*}, \label{Eqn: TDMA Proof1}
\end{eqnarray}
since any set of received powers that can be achieved under $\vec{h}$ can be achieved under $\vec{g}$. Now, note that if $\pc^*$ schedules only one user for transmission with full power at $\vec{g}$, then it schedules only the best user for transmission with full power at $\vec{h}$ since the maximum sum-rate at $\vec{g}$ forms an achievable upper bound for the maximum sum-rate at $\vec{h}$ for this case.

By using the structural properties of $\pc^*$ established in Theorem \ref{Thm: Optimum Power Control 1}, we can write $R_{\vec{g}}\paren{\pc^*}$ as
\begin{eqnarray}
R_{\vec{g}}\paren{\pc^*} &=& \frac12 \sum_{i=1}^{k^*} \log\paren{1+\frac{\imax{h}{1}}{\rho^{-1} + \paren{k^* - 1}\imax{h}{1}}} \nonumber \\
&=& \frac12 k^* \log\paren{1+\frac{\rho \imax{h}{1}}{1 + \paren{k^* - 1} \rho \imax{h}{1}}} \nonumber
\end{eqnarray}
for some optimal $k^* \in \brparen{1, \cdots, n}$. Our aim is to find a condition on $\imax{h}{1}$ under which we can show that $k^* = 1$.

A similar problem was addressed in \cite{Hanly10} but for a different model: the symmetric network of interfering links. This is a model in which there are $n$ links, each with a different receiver node, and each link interferes with all the others. The symmetry refers to the fact that the direct link gain is unity for all links, and the cross-link gain is $\sqrt{\epsilon}$ between any pair of links. See figure 1 in \cite{Hanly10} for an illustration of this model. In \cite{Hanly10} the received power is denoted by $P_{max}$ but if we replace that by $\rho \imax{h}{1}$ then the sum-rate in this model, with $n$ links on, is given by
\begin{equation}
R_n(\epsilon) = n \log\paren{1+\frac{\rho \imax{h}{1}}{1 + \epsilon \paren{n - 1} \rho \imax{h}{1}}}. \nonumber
\end{equation}
Note that if $\epsilon = 1$ then this gives the same rate as $n$ links on in the model of the present appendix, under fading state $\vec{g}$, and indeed the symmetric network model degenerates into, effectively, a symmetric multiple access model in the special case $\epsilon = 1$.

We can use results from \cite{Hanly10}, Section IV B, to obtain the condition on $\imax{h}{1}$ that we need. Section IV B examines the special case of binary power control in which a link is either on at full power or switched right off.  First, it is shown that $R_n(\epsilon)$ is a decreasing function of $\epsilon$, and it crosses the constant value $R_1$ at a unique value of $\epsilon$, namely,
\begin{equation}
\epsilon_{n,1} = \frac{(1+\rho \imax{h}{1}) -
(1+\rho \imax{h}{1})^{\frac{1}{n}}}{(n-1)\rho \imax{h}{1} ((1+\rho \imax{h}{1})^
                    {\frac{1}{n}} - 1)}
                    \label{eq:en1}
\end{equation}
(see (36) in \cite{Hanly10}). Thus, if $\epsilon > \epsilon_{n,1}$, then having one link on beats having $n$ links on. Further, it is shown in Lemma 4.3 in \cite{Hanly10} that $\epsilon_{n,1}$ increases in $n$, and approaches a limiting value of $\epsilon^* := \displaystyle (\log(1 + \rho \imax{h}{1}))^{-1}$ as $n$ tends to infinity. Thus, if $\epsilon > \epsilon^*$, having one link on must be optimal in the class of binary power control schemes.

If we can show that $1 > \epsilon^*$ then it will follow that having one link on is optimal in our multiple access model under fading $\vec{g}$. But if $\imax{h}{1} > (e-1) \rho^{-1}$ then indeed $1 > \epsilon^*$, so we conclude that a sufficient condition for scheduling just the best link is $\imax{h}{1} > (e-1) \rho^{-1}$, as stated in the theorem. 

{}


\begin{thebibliography}{}

\bibitem{WSJ10}
R. Bender and G. Sandstrom, ``Wireless carriers refine 4G technology," {\em Wall Street Journal}, Available Online: http://online.wsj.com/article/SB10001424052748704869304575109624056054264.html, March 2010.

\bibitem{ITU08}
International Telecommunication Union, ``Requirements related to technical performance for IMT-Advanced radio interface(s)," {\em ITU-R M.2134 Technical Report}, Available Online: http://www.itu.int/publ/R-REP-M.2134-2008/en, Nov. 2008.

\bibitem{Tse09}
D. N. C. Tse, ``Interference management: an information theoretic view," {\em 2009 IEEE International Symposium on Information Theory}, Tutorial T4, Available Online: http://www.eecs.berkeley.edu/~dtse, June 2009.

\bibitem{HT99}
S. Hanly and D. Tse,``Power Control and Capacity of Spread Spectrum Wireless Networks", {\em Automatica}, Vol. 35, No. 12, pp. 1987-2012, Dec. 1999.

\bibitem{Zhang08}
Z. Luo and S. Zhang, ``Dynamic spectrum management: complexity and duality," {\em IEEE Journal of Selected Topics in Signal Processing}, vol. 2, no. 1, pp. 57-73, Feb. 2008.

\bibitem{Gesbert08}
A. Gjendemsj\o, D. Gesbert, G. E. \O{ien} and S. G. Kiani, ``Binary power control for sum rate maximization over multiple interfering links," {\em IEEE Trans. Wireless Commun.,} vol. 7, no. 8, pp. 3164-3173, Aug. 2008.

\bibitem{JG03}
S. A. Jafar and A. Goldsmith, ``Adaptive multirate CDMA for uplink throughput maximization," {\em IEEE Trans. Wireless Commun.,} vol. 2, no. 2, March 2003.

\bibitem{OZW03}
S.-J. Oh, D. Zhang and K. M. Wasserman, ``Optimal resource allocation in multiservice CDMA networks," {\em IEEE Trans. Wireless Commun.,} vol. 2, no. 4, pp. 811-821, July 2003.

\bibitem{Oh06}
S.-J. Oh and A. C. K. Soong, ``QoS-constrained information-theoretic sum capacity of reverse link CDMA systems," {\em IEEE Trans. Wireless Commun.,} vol. 5, no. 1, pp. 3-7, Jan. 2006.

\bibitem{Hanly10}
S. R. Bhaskaran, S. V. Hanly, N. Badruddin and J. S. Evans, ``Maximizing the sum rate in symmetric networks of interfereing links," {\em IEEE Trans. Inform. Theory}, to appear.

\bibitem{Palomar06}
D. P. Palomar and Y. Jiang, ``MIMO transceiver design via majorization theory," {\em Foundations and Trends in Communications and Information Theory, Now Publishers}, vol. 3, no. 4-5, pp. 331-551, 2006.

\bibitem{Viswanath99a}
P. Viswanath and V. Anantharam, ``Optimal sequences and sum capacity of synchronous CDMA systems," {\em IEEE Trans. Inform. Theory}, vol. 45, no. 6, pp. 1984-1991, Sept. 1999.

\bibitem{Viswanath02}
P. Viswanath and V. Anantharam, ``Optimal sequences for CDMA under colored noise: A Schur-saddle function property," {\em IEEE Trans. Inform. Theory}, vol. 48, no. 6, pp. 1295-1318, June 2002.

\bibitem{Viswanath99b}
P. Viswanath, V. Anantharam and D. N. C. Tse, ``Optimal sequences, power control, and user capacity of synchronous CDMA systems with linear MMSE multiuser receivers," {\em IEEE Trans. Inform. Theory}, vol. 45, no. 6, pp. 1968-1983, Sept. 1999.

\bibitem{Knopp95}
R. Knopp and P. A. Humblet, ``Information capacity and power control in single cell multiuser communications," {\em Proceedings of the International Conference on Communications}, Seattle, WA, June 1995.

\bibitem{RU96}
B. Rimoldi and R. Urbanke, ``A rate-splitting approach to the Gaussian multiple-access channel," {\em IEEE Trans. Inform. Theory}, vol. 42, no. 2, pp. 364-375, March 1996.

\bibitem{HanlyTse98a}
D. N. C. Tse and S. V. Hanly, ``Multiaccess fading channels - part I: polymatroid structure, optimal resource allocation and throughput capacities," {\em IEEE Trans. Inform. Theory}, vol. 44, no. 7, pp. 2796-2815, Nov. 1998.

\bibitem{SZZ08}
S. Sambhwani, W. Zhang and W. Zei, ``Uplink interference cancellation in HSPA: principles and practice," {\em Available online: http://www.qualcomm.com/common/documents/white\_papers/ul-ic-hspa.pdf}.

\bibitem{Arnold07}
B. C. Arnold, ``Majorization: Here, there and everywhere," {\em Statist. Sci.}, vol. 22, no. 3, pp. 407-413, 2007.

\bibitem{Olkin79}
A. W. Marshall and I. Olkin, {\it Inequalities: Theory of Majorization and Its Applications}, Academic Press, New York, 1979

\bibitem{Arnold87}
B. C. Arnold, \textit{Majorization and the Lorenz Order: A Brief Introduction}, Springer-Verlag, New York, 1987.

\bibitem{CT06}
T. M. Cover and J. A. Thomas, {\it Elements of Information Theory}, Wiley, New York, 2nd ed., July 2006.

\bibitem{ICPW09}
H. Inaltekin, M. Chiang, H. V. Poor and S. B. Wicker, ``On unbounded path-loss models: effects of singularity on wireless network performance," {\em IEEE J. Select. Areas Commun.}, vol. 27, no. 7, pp. 1078-1092, Sept. 2009.

\bibitem{Viterbi90}
A. J. Viterbi, ``Very low rate convolutional codes for maximum theoretical performance of spread-spectrum multiple-access channels," {\em IEEE J. Select. Areas Commun.}, vol. 8, no. 4, pp. 641-649, May 1990.

\bibitem{HSPT06}
J. Hou, J. E. Smee, H. D. Pfister and S. Tomasin, ``Implementing interference cancellation to increase the EV-DO Rev A reverse link capacity," {\em IEEE Communications Magazine}, vol. 44, no. 2, pp. 58-64, Feb. 2006.

\bibitem{HW93} S.V. Hanly and P. Whiting, ``Information-theoretic capacity of multi-receiver networks," {\em Telecommunication Systems}, vol. 1, no. 1, pp.1-42, 1993.



\end{thebibliography}
\end{document}